\documentclass[letter,longauth]{aa}  

\usepackage{graphicx}
\usepackage{gensymb}

\usepackage{txfonts}
\usepackage[flushleft]{threeparttable}
\renewcommand{\arraystretch}{1.1}
\usepackage[hidelinks,colorlinks=true,linkcolor=blue,citecolor=blue]{hyperref}
\usepackage{xcolor}
\usepackage{longtable}
\usepackage{lipsum}
\usepackage{adjustbox}
\usepackage{stfloats}
\usepackage{scrextend}
\usepackage{amssymb}
\usepackage{amsmath}
\usepackage{mathrsfs}
\usepackage{multicol}
\usepackage{booktabs}
\usepackage{moresize}
\usepackage{orcidlink}
\usepackage{xspace}
\usepackage{placeins}
\usepackage{multirow}
\usepackage{booktabs}
\usepackage{diagbox}

\newcommand{\teff}{\ensuremath{T_{\rm eff}}}

\newcommand{\kms}{\mbox{km\,s$^{-1}$}\xspace}
\newcommand{\ms}{\mbox{m\,s$^{-1}$}\xspace}

\newcommand{\mplanet}{\mbox{$M_{\rm p}$}\xspace}
\newcommand{\rplanet}{\mbox{$R_{\rm p}$}\xspace}
\newcommand{\mjup}{\mbox{$\it{M_{\rm \mathrm{Jup}}}$}\xspace}

\newcommand{\me}{\mbox{$\it{M_{\rm \mathrm{\oplus}}}$}\xspace}
\newcommand{\re}{\mbox{$\it{R_{\rm \mathrm{\oplus}}}$}\xspace}
\newcommand{\mstar}{\mbox{$M_{*}$}\xspace}
\newcommand{\rstar}{\mbox{$R_{*}$}\xspace}

\newcommand{\rsol}{\mbox{$\it{R_\mathrm{\odot}}$}\xspace}

\begin{document}

   \title{Upside down: GJ\,3090\,b the first retrograde exoplanet
around an M dwarf detected with NIRPS}

\titlerunning{GJ\,3090\,b a retrograde exoplanet
around an M dwarf}
\authorrunning{Y. Carteret, et al.}
   \author{Yann Carteret\inst{1}
\and Vincent Bourrier\inst{1}
\and L\'ena Parc\inst{1}
\and Andrew Winter\inst{2}
\and Romain Allart\inst{3}
\and Izan de Castro Le\~ao\inst{4}
\and Fran\c{c}ois Bouchy\inst{1}
\and Vera M. Passegger\inst{5,6,7,8}
\and Charles Cadieux\inst{1,3}
\and Pierrot Lamontagne\inst{9,3}
\and \'Etienne Artigau\inst{3,10}
\and Fr\'ed\'erique Baron\inst{3,10}
\and Susana C. C. Barros\inst{11,12}
\and Bj\"orn Benneke\inst{13,3}
\and Xavier Bonfils\inst{14}
\and Marta Bryan\inst{15}
\and Bruno L. Canto Martins\inst{4}
\and Ryan Cloutier\inst{16}
\and Eduardo Cristo\inst{11}
\and Jonathan Gagn\'e\inst{17,3}
\and Neil J. Cook\inst{3}
\and Nicolas B. Cowan\inst{18,19}
\and Jose Renan De Medeiros\inst{4}
\and Xavier Delfosse\inst{14}
\and Ren\'e Doyon\inst{3,10}
\and Xavier Dumusque\inst{1}
\and David Ehrenreich\inst{1,20}
\and Jonay I. Gonz\'alez Hern\'andez\inst{5,6}
\and David Lafreni\`ere\inst{3}
\and Christophe Lovis\inst{1}
\and Lison Malo\inst{3,10}
\and Claudio Melo\inst{21}
\and Lucile Mignon\inst{1,14}
\and Christoph Mordasini\inst{22}
\and Francesco Pepe\inst{1}
\and Rafael Rebolo\inst{5,6,23}
\and Jason Rowe\inst{24}
\and Nuno C. Santos\inst{11,12}
\and Damien S\'egransan\inst{1}
\and Alejandro Su\'arez Mascare\~no\inst{5,6}
\and St\'ephane Udry\inst{1}
\and Diana Valencia\inst{15}
\and Gregg Wade\inst{25,26}
\and Khaled Al Moulla\inst{11,1}
\and Rillck Guilherme de Souza Barros de Amorim\inst{4}
\and Lisa Dang\inst{27}
\and Emily Deibert\inst{27}
\and Dasaev O. Fontinele\inst{4}
\and Thierry Forveille\inst{14}
\and Yolanda G. C. Frensch\inst{11,1}
\and Roseane de Lima Gomes\inst{3,4}
\and Dany Mounzer\inst{1}
\and Stefan Pelletier\inst{1,3}
\and Riley Rosener\inst{3}
\and Bennett Neil Skinner\inst{16,28}
\and Avidaan Srivastava\inst{3}
\and Atanas K. Stefanov\inst{5,6}
\and Valentina Vaulato\inst{1}
\and Joost P. Wardenier\inst{22,3}
\and Drew Weisserman\inst{16}}

   \institute{
\inst{1}Observatoire de Gen\`eve, D\'epartement d’Astronomie, Universit\'e de Gen\`eve, Chemin Pegasi 51, 1290 Versoix, Switzerland\\
\inst{2}Astronomy Unit, School of Physics and Astronomy, Queen Mary University of London, London E1 4NS, UK\\
\inst{3}Institut Trottier de recherche sur les exoplan\`etes, D\'epartement de Physique, Universit\'e de Montr\'eal, Montr\'eal, Qu\'ebec, Canada\\
\inst{4}Departamento de F\'isica Te\'orica e Experimental, Universidade Federal do Rio Grande do Norte, Campus Universit\'ario, Natal, RN, 59072-970, Brazil\\
\inst{5}Instituto de Astrof\'isica de Canarias (IAC), Calle V\'ia L\'actea s/n, 38205 La Laguna, Tenerife, Spain\\
\inst{6}Departamento de Astrof\'isica, Universidad de La Laguna (ULL), 38206 La Laguna, Tenerife, Spain\\
\inst{7}Hamburger Sternwarte, Gojenbergsweg 112, D-21029 Hamburg, Germany\\
\inst{8}Subaru Telescope, National Astronomical Observatory of Japan (NAOJ), 650 N Aohoku Place, Hilo, HI 96720, USA\\
\inst{9}Instituto de Astrof\'isica de Andaluc\'ia (IAA-CSIC), Glorieta de la Astronom\'ia s/n, 18008, Granada, Spain\\
\inst{10}Observatoire du Mont-M\'egantic, Qu\'ebec, Canada\\
\inst{11}Instituto de Astrof\'isica e Ci\^encias do Espa\c{c}o, Universidade do Porto, CAUP, Rua das Estrelas, 4150-762 Porto, Portugal\\
\inst{12}Departamento de F\'isica e Astronomia, Faculdade de Ci\^encias, Universidade do Porto, Rua do Campo Alegre, 4169-007 Porto, Portugal\\
\inst{13}Department of Earth, Planetary, and Space Sciences, University of California, Los Angeles, CA 90095, USA\\
\inst{14}Univ. Grenoble Alpes, CNRS, IPAG, F-38000 Grenoble, France\\
\inst{15}Department of Physics, University of Toronto, Toronto, ON M5S 3H4, Canada\\
\inst{16}Department of Physics \& Astronomy, McMaster University, 1280 Main St W, Hamilton, ON, L8S 4L8, Canada\\
\inst{17}Plan\'etarium de Montr\'eal, Espace pour la Vie, 4801 av. Pierre-de Coubertin, Montr\'eal, Qu\'ebec, Canada\\
\inst{18}Department of Physics, McGill University, 3600 rue University, Montr\'eal, QC, H3A 2T8, Canada\\
\inst{19}Department of Earth \& Planetary Sciences, McGill University, 3450 rue University, Montr\'eal, QC, H3A 0E8, Canada\\
\inst{20}Centre Vie dans l’Univers, Facult\'e des sciences de l’Universit\'e de Gen\`eve, Quai Ernest-Ansermet 30, 1205 Geneva, Switzerland\\
\inst{21}European Southern Observatory (ESO), Karl-Schwarzschild-Str. 2, 85748 Garching bei M\"unchen, Germany\\
\inst{22}Space Research and Planetary Sciences, Physics Institute, University of Bern, Gesellschaftsstrasse 6, 3012 Bern, Switzerland\\
\inst{23}Consejo Superior de Investigaciones Cient\'ificas (CSIC), E-28006 Madrid, Spain\\
\inst{24}Bishop's University, Dept of Physics and Astronomy, Johnson-104E, 2600 College Street, Sherbrooke, QC, Canada, J1M 1Z7, Canada\\
\inst{25}Department of Physics, Engineering Physics, and Astronomy, Queen’s University, 99 University Avenue, Kingston, ON K7L 3N6, Canada\\
\inst{26}Department of Physics and Space Science, Royal Military College of Canada, 13 General Crerar Cres., Kingston, ON K7P 2M3, Canada\\
\inst{27}Department of Physics and Astronomy, University of Waterloo, 200 University W, Waterloo, ON N2L 3G1, Canada\\
\inst{28}Origins Institute, McMaster University, 1280 Main St W, Hamilton, ON, L8S 4L8, Canada\\
}

   \date{Received 14 July, 2026; accepted 15 August, 2026}

  \abstract{The angle between stellar spin axis and planetary orbital plane can provide key insights into the formation and dynamical evolution of planetary systems. In particular, this measurement in multi-planet systems can be used to further discriminate between different competing migration scenarios. We present six transit observations of the sub-Neptune GJ\,3090\,b obtained with NIRPS and HARPS. GJ\,3090\,b is the inner planet of a confirmed multi-planet system orbiting an M dwarf (M2). Using high spectral resolution and high temporal cadence spectroscopic observations, we analyzed the Rossiter-McLaughlin (RM) effect induced by GJ\,3090\,b to determine its orbital obliquity. Through the RM revolutions technique, we find that the planet is on a retrograde orbit with a derived 3D obliquity of $\psi = 136^{+24}_{-18}\,\mathrm{deg}$. We find no evidence of massive outer planetary or wide stellar binary companions, which disfavors scenarios involving gravitational perturbations from a massive body and instead points toward a primordial misalignment of the protoplanetary disk. Our results establish GJ\,3090\,b as the first planet on a retrograde orbit discovered around an M dwarf and the first highly misaligned confirmed multi-planet system without a known massive companion. We further propose late secondary disk accretion around GJ\,3090, in which the disk is not expected to be aligned with the stellar spin axis, followed by disk-driven migration as the most likely mechanisms to explain the observed architecture. This work also illustrates the capability of the RM revolutions technique when applied to near-infrared data to probe orbital architectures of the smallest planets around M dwarfs that have remained mostly inaccessible. 
  }


   \maketitle
%

\section{Introduction}
Orbital architecture measurements of planetary systems provide key information on their formation and dynamical evolution and help constrain their different migration pathways. Planets are expected to form within a protoplanetary disk and inherit its angular momentum, resulting in prograde orbits aligned with the rotation axis of their host star \citep[e.g.,][]{Winn2015}. However, a variety of mechanisms acting at different stages of its evolution can alter the expected primordial alignment of the system. During the system's formation, a tilt of the protoplanetary disk, driven for instance by the gravitational influence of a misaligned binary companion \citep[see, e.g.,][]{Lai2014} or a tilt of the stellar spin axis \citep[e.g., torques from stellar winds;][]{Spalding2019} can trigger an initial misalignment. At later stages, gravitational interactions between the different planets of the system can occasionally send planets on highly misaligned orbits \citep[see][]{Wu2003,Chatterjee2008,Teyssandier2013}. Multi-planetary systems offer a unique opportunity to disentangle these competing channels, as the mutual inclination and dynamical stability of the orbits can be used to discriminate between quiescent disk migration and dynamically disruptive processes. Counteracting these excitation mechanisms, stellar tides are expected to dampen obliquities after inward migration by progressively realigning the orbits of close-in planets \citep{Lai2012}. This tidal realignment operates most efficiently around stars with deep convective envelopes, such as M dwarfs \citep{Winn2010,Attia2023}. In this context, multi-planetary systems around M dwarfs are ideal candidates to probe the competition between tidal damping and dynamical excitation.

Analysis of time series stellar spectra at a high spectral resolution can be used to probe the local stellar surface occulted by the planet along its transit chord \citep[][]{Triaud2018}. The rotation of the stellar photosphere causes one limb of the star to be blueshifted and the other to be redshifted. As the planet transits, it will successively occult stellar regions with different line-of-sight velocities, affecting the observed disk-integrated stellar lines. The resulting distortion, also called the Rossiter-McLaughlin (RM) effect (\citealt{Rossiter1924,McLaughlin1924}), shifts the measured centroid of the stellar lines from the expected Keplerian curve. The shape and amplitude of this radial velocity (RV) anomaly encode the sky-projected spin-orbit angle ($\lambda$) and offer access to the orbital architecture of the system.

In this study, we focus on the close-in sub-Neptune GJ\,3090\,b ($\rplanet=2.18$\,\re, $\mplanet=4.52$\,\me) orbiting an M dwarf with a period of approximately $2.9\,\mathrm{days}$ \citep{Almenara2022}. The system also hosts an additional candidate sub-Neptune at 13\,days and a confirmed non-transiting sub-Neptune at 16\,days \citep{Lamontagne2026}. In this letter, we present the first measurement of GJ\,3090\,b's orbital configuration, which we use to constrain the dynamical history of this system and the migration pathways that led to its current architecture.

\vspace{-0.3cm}

\section{Observations}

We present observations of the M dwarf GJ\,3090 ($m_V=11.4$, $m_J=8.2$) obtained as part of the NIRPS guaranteed time observation (GTO) sub-package 3 \citep[][program ID: 112.25P3]{Allart2025,NIRPS} with the 3.6 m ESO telescope at La Silla, Chile. We observed six transits of GJ\,3090\,b with NIRPS \citep{NIRPS} and four simultaneously with HARPS \citep{HARPS}. Together NIRPS and HARPS cover a wide wavelength range, from optical (380–690 nm) to near-infrared (980–1800 nm) at high spectral resolution. We collected a total of 142 spectra with NIRPS and 51 with HARPS covering a total of about 23 hours of observations. Further details about the observations are listed in Table~\ref{tab:observations}. 
\vspace{-0.3cm}
\begin{figure}[ht]
    \centering
    \includegraphics[width=0.8\columnwidth]{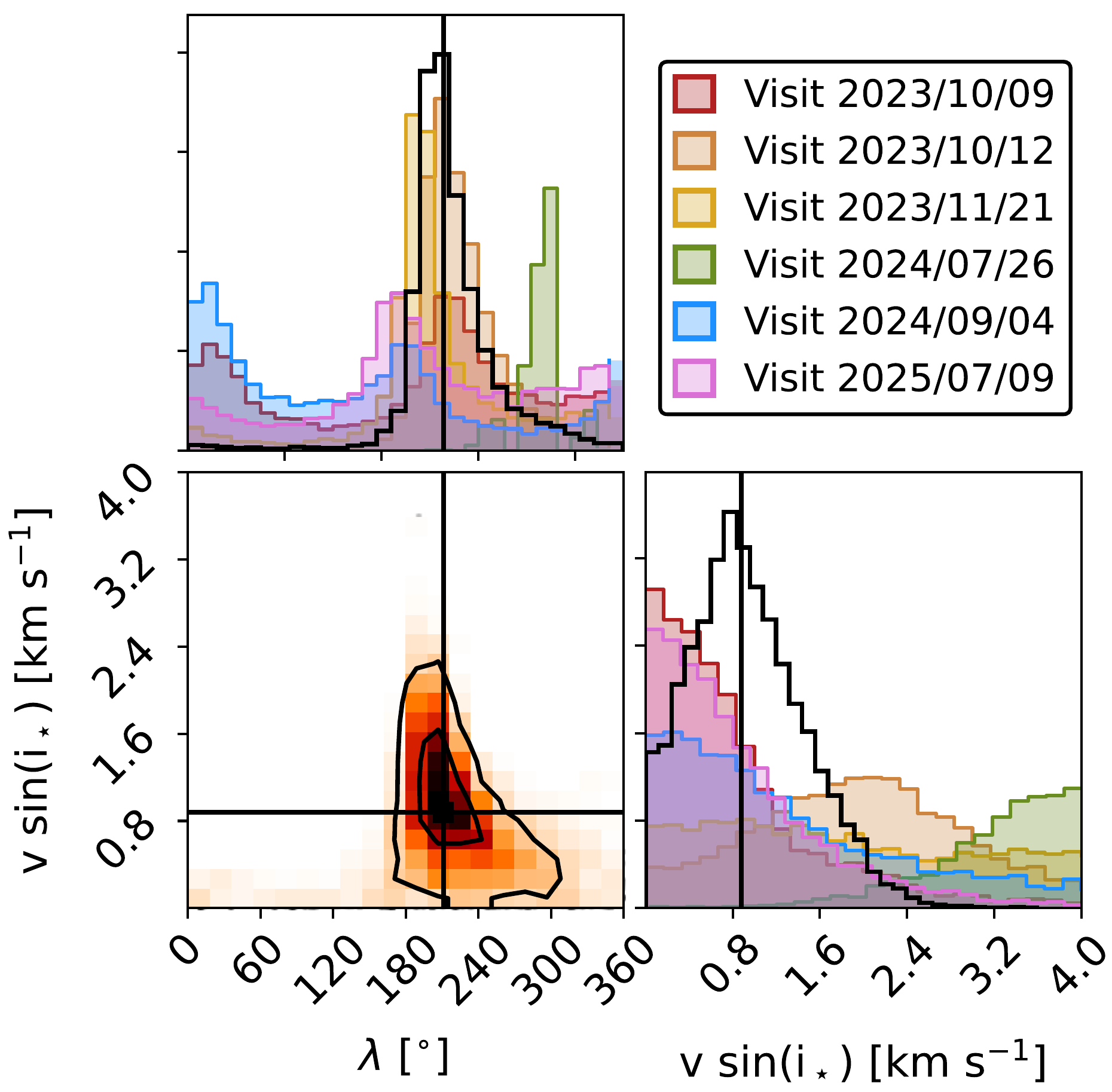}
    \caption{Posterior distributions of projected obliquity and stellar rotation velocity for NIRPS. Fits performed on individual nights are displayed by colored shading, while the combined fit of the five NIRPS (excluding visit 4, green PDF) transits are shown by the black PDF. The black line denotes the median values of the distributions. }
    \label{fig:lambda_veq}
\end{figure}
\vspace{-0.3cm}

We performed the data reduction using the \texttt{\href{https://gitlab.unige.ch/spice_dune/antaress}{ANTARESS}} pipeline \citep{Bourrier2024}. The 2D (order-separated) echelle spectra that \texttt{ANTARESS} takes as input were obtained using the Data Reduction Software (DRS) of NIRPS v3.3.12 and HARPS v3.3.6, both adapted from the ESPRESSO pipeline \citep{Pepe2021}. In particular, we refined the telluric absorption correction (based on \citealt{Allart2022}) for our dataset by adapting the line list over which water and methane tellurics are fit. Using the out-of-transit observed spectra, we further computed an optimized custom cross-correlation function (CCF) mask to extract RVs. More information about the reduction steps can be found in Appendix~\ref{appdx:reduction_steps}.

\vspace{-0.3cm}
\section{Rossiter-McLaughlin analysis}
\label{sec:RM_analysis}

We modeled the RM effect using the RM revolutions (RMR; \citealt{Bourrier2021}) approach implemented in \texttt{ANTARESS}. As a first step, we fit each visit and instrument independently to ensure consistency across our visits. The individual resulting probability distribution functions (PDFs) for NIRPS and HARPS are shown in Fig.~\ref{fig:lambda_veq} and Fig.~\ref{fig:lambda_veq_HARPS}. The NIRPS observations provide tight constraints on $\lambda$, with consistent PDFs across individual visits except for the 26 July 2024 night (hereafter visit 4), which has the lowest and most variable signal-to-noise ratio (S/N) during the transit across our entire dataset. It is also the only visit for which the Bayesian information criterion (BIC) favored the null hypothesis over the model. In contrast, HARPS transits weakly constrain the projected stellar rotational velocity and $\lambda$, which is explained by the coarse temporal sampling (Table~\ref{tab:observations}), the reduced number of visits, and the lower S/N in the visible compared to the near-infrared for an M dwarf. We thus discarded NIRPS visit 4 and the HARPS data for the final joint fit (see also Appendix~\ref{appdx:RM_fit}).

\begin{figure}[ht]
    \centering
    \includegraphics[width=0.9\columnwidth]{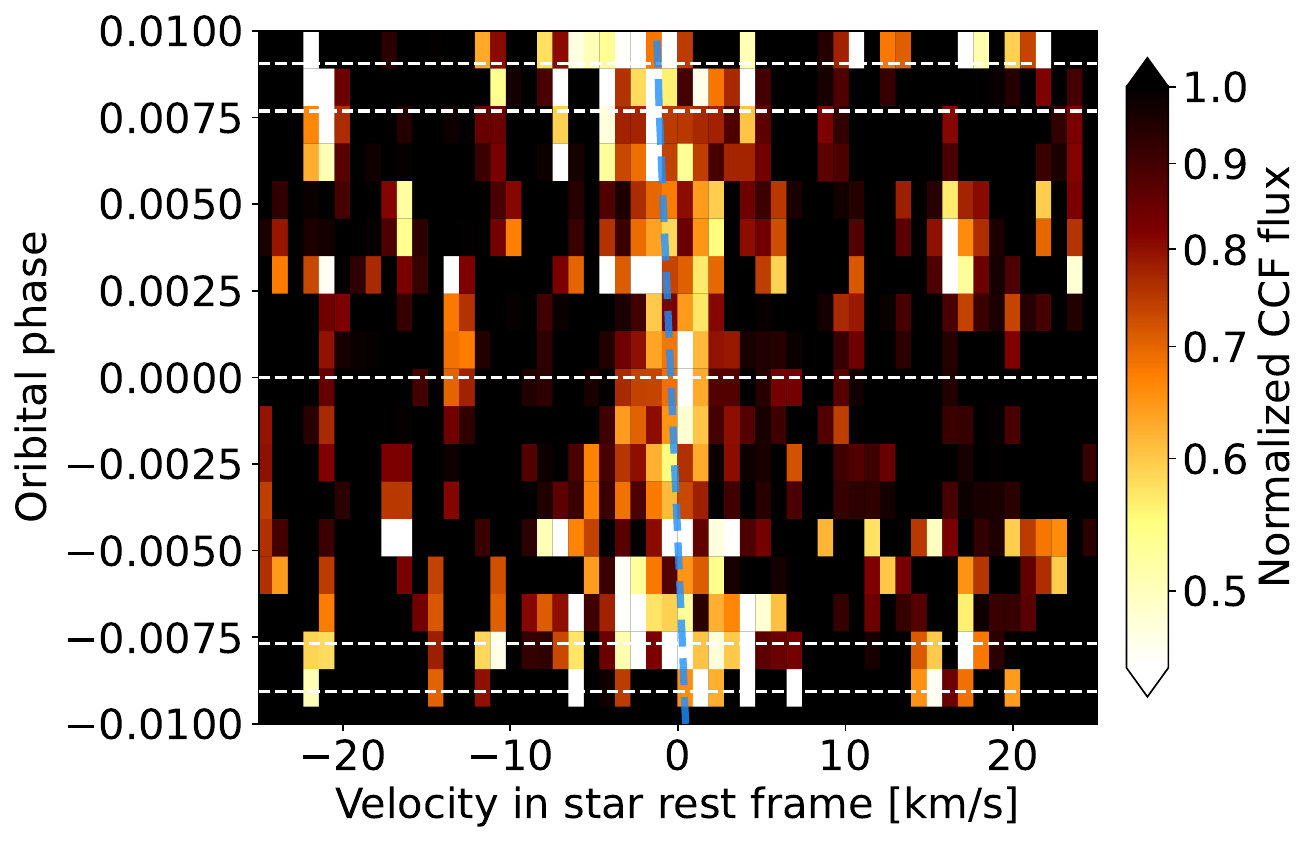}
    \caption{Intrinsic CCF map averaged over the five retained visits of NIRPS. Dashed horizontal lines denote the contact points and mid-transit. The dashed blue line corresponds to the RVs' best-fit model local stellar lines occulted by the planet.}
    \label{fig:intr_maps}
\end{figure}

\begin{figure}[ht]
    \centering
    \includegraphics[width=0.65\columnwidth]{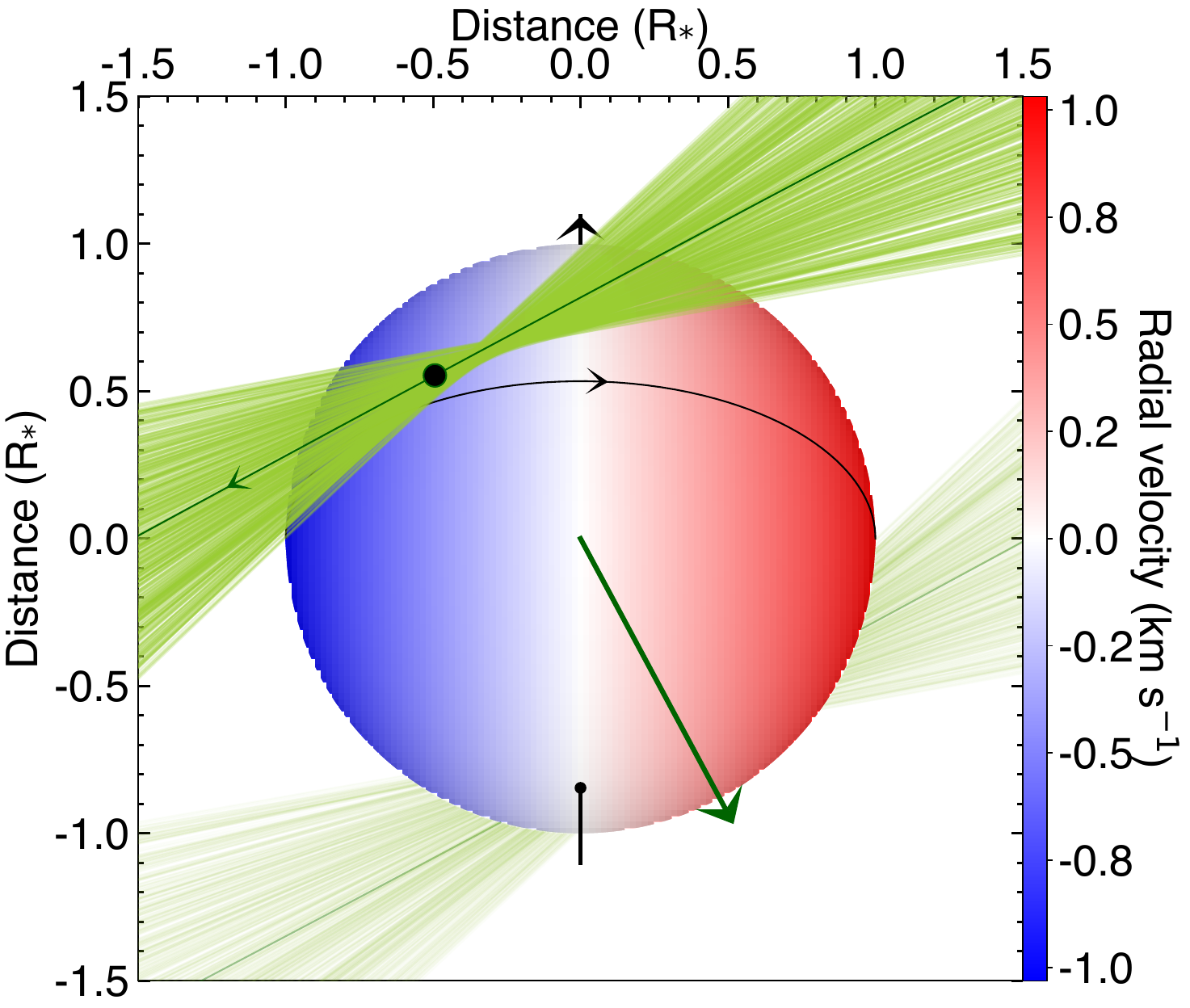}
    \caption{Best-fit architecture of the GJ\,3090 system, adopting the south $i_*$ configuration. The dark green and black arrows refer to, respectively, the planetary orbital and the stellar rotation planes. The light green lines represent the possible orbital trajectory of GJ\,3090\,b, sampled within $1-\sigma$ from the posterior distribution of $\lambda$. The stellar surface is colored as a function of the photospheric RV field.}
    \label{fig:syst_view}
\end{figure}

Finally, we performed a joint fit over the five remaining NIRPS visits for $\lambda$, the stellar rotation period ($P_\mathrm{rot}$), the stellar radius ($\rstar$), and the stellar inclination ($i_*$), adopting priors on the stellar and planetary parameters from \citet{Lamontagne2026}. The resulting posterior distributions are shown in Fig.~\ref{fig:full_corner_psi}, and the intrinsic (planet occulted) CCF map with the best-fit model is presented in Fig.~\ref{fig:intr_maps}. The combined Bayesian modeling of the RM effect yields a projected obliquity of $\lambda = 208^{+15}_{-18}\,\mathrm{deg}$, from which we computed the 3D orbital obliquity $\psi$. Combining the posterior distributions of $P_\mathrm{rot}$ and the planetary and stellar (see Table~\ref{tab:RM_fits}) inclinations, we derived $\psi = 136^{+24}_{-18}\,\mathrm{deg}$\footnote{The derived posterior on $i_*$ is almost symmetric around 90\,deg, lifting the northern and southern configurations' degeneracy of $\psi$.} for GJ\,3090\,b, indicative of a retrograde orbit. A summary of the priors used and posteriors derived can be found in Appendix~\ref{appdx:RM_fit}, and the system's architecture is shown in Fig.~\ref{fig:syst_view}. We note that the small size of the planet and the moderate stellar rotation velocity prevent the detection of the classical and reloaded \citep{Cegla2016} RM anomaly in our dataset (see Fig.~\ref{fig:classical_RM}). Notably, GJ\,3090\,b becomes the smallest planet around an M dwarf with a 3D obliquity measurement. This result, being fully driven by NIRPS, highlights that near-infrared high-resolution spectroscopy from a stabilized spectrograph, processed with state-of-the-art tools such as \texttt{ANTARESS} and analyzed with the RMR technique, can enable obliquity measurements of small planets around cool stars.

\section{Discussion}

\subsection{A sub-Neptune on a retrograde orbit around an M dwarf}

\begin{figure}[ht]
    \centering
    \includegraphics[width=0.75\columnwidth]{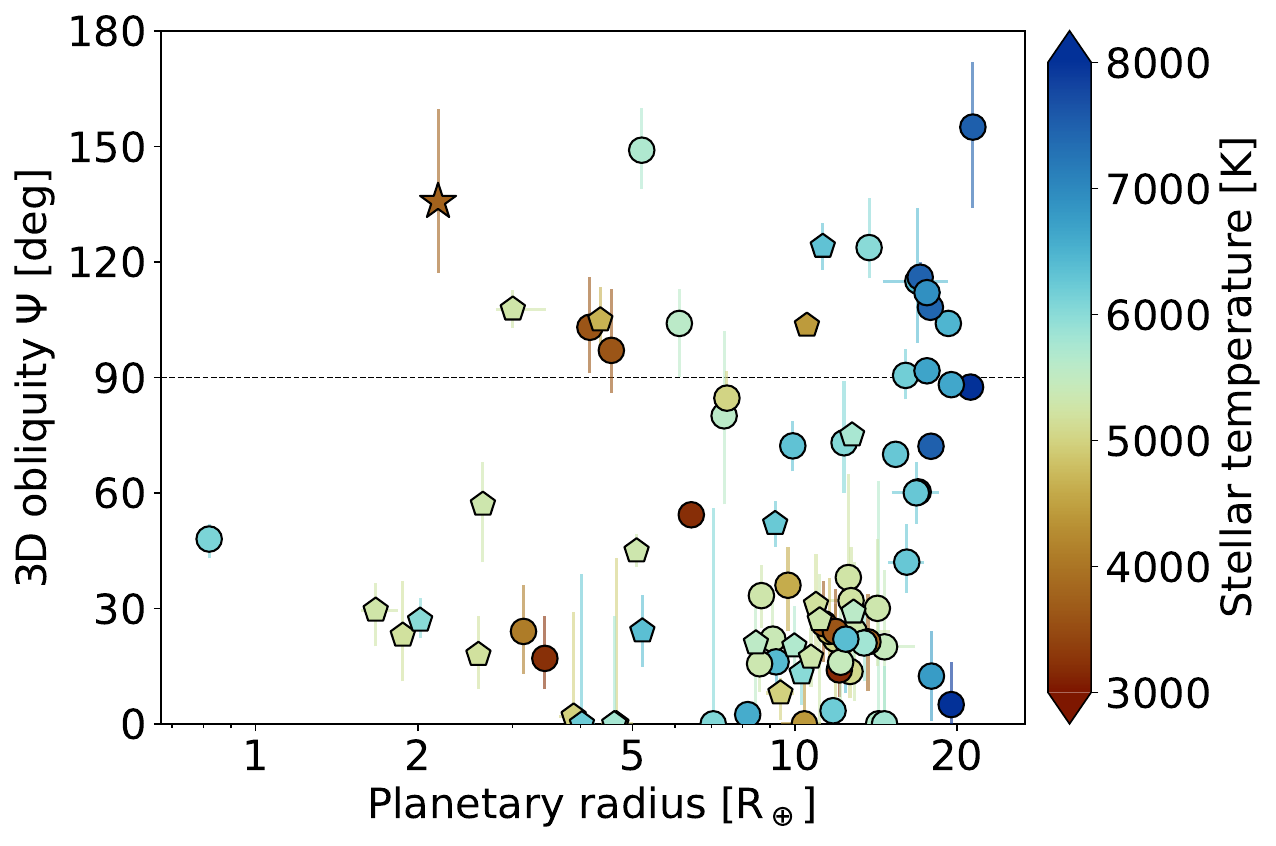}
    \vspace{-0.2cm}
    \caption{Measured 3D obliquity of exoplanets compared to their size from TEPCat and the NASA Exoplanet Archive as of July 2026. GJ\,3090\,b is denoted by the star, while pentagon markers indicate planets in a multi-system. All planets are colored as a function of their host temperature.}
    \label{fig:psi_pop}
\end{figure}

The derived 3D obliquity measurement makes GJ\,3090\,b one of the most misaligned planets (together with the two giants KELT-19\,b \citealt{Kawai2024} and TOI-1710\,b \citealt{Mantovan2026}) among those for which a $\psi$ measurement has been reported (TEPCat, \citealt{Southworth2026}). In Fig.~\ref{fig:psi_pop}, GJ\,3090\,b is located above the group of polar planets ($75<\psi<115$) that were previously identified in \citet{Albrecht2021} and \citet{Attia2023} and argued to be stabilized by tides. Due to their large convective envelopes, tides around M dwarfs are particularly efficient at realigning planetary orbits. However, because this low-mass planet is barely tidally detached ($a/\rstar\sim13>10$, \citealt{Rice2021}), the tidal efficiency factor of GJ\,3090\,b, computed using Eq.~4 of \citet{Attia2023}, falls on the lower half of the planets with a measured $\lambda$, which is consistent with a weak tidal realignment. Moreover, we computed an age of $1.07_{-0.16}^{+0.71}\,\mathrm{Gyr}$ for the system (see Appendix~\ref{appdx:system_properties}). This sets GJ\,3090\,b just at the edge of the trend noted in \citet{Mantovan2024} for planets around cool stars (defined from the Kraft break $\teff<6250\,\mathrm{K}$; \citealt{Kraft1967}) in which highly misaligned systems are old ($>1$\,Gyr) and tidally detached. Among M dwarf hosts, GJ\,3090\,b stands as the smallest planet characterized and the first retrograde system.

GJ\,3090 is one of the six confirmed multi-planetary systems hosting a highly misaligned planet ($\psi>70^\circ$). The prevailing explanation for such strong misalignment involves mechanisms driven by the gravitational influence of a massive companion acting after the dissipation of the protoplanetary disk. These include high-eccentricity tidal migration (HEM) --- via planet-planet scattering (e.g., \citealt{Ford2008}) or Kozai-Lidov cycles (e.g., \citealt{Wu2003}) --- and secular interaction \citep{Boue2014}, which can be triggered either by an outer massive planetary companion \citep[e.g., for WASP-107, HAT-P-11, and HD\,118203;][]{Bourrier2023,Knudstrup2024} or by the presence of a wide binary stellar companion \citep[e.g., for WASP-131 and KELT-19,][]{Doyle2023,Kawai2024}. Consistently with this picture, all five previously known highly misaligned multi-planetary systems have either a confirmed wide binary stellar companion or a massive outer planetary companion, whereas no such perturber has been identified in GJ\,3090.

\vspace{-0.8cm}

\subsection{The search for companions}

We searched for additional planetary, brown dwarf, or stellar companions in the system by combining radial velocities, high angular resolution imaging, and Gaia DR3 photometry. GJ\,3090 has been extensively monitored with HARPS and NIRPS as part of the RV campaign previously published in \citet{Lamontagne2026}. We used the residual RVs from that study to place limits on the mass and orbital separation of a possible outer companion (see Fig.~\ref{fig:companion_mass_limit}). Details of our methodology are given in Appendix~\ref{appdx:system_properties}. By combining the RV, speckle-imaging, and Gaia constraints, we could exclude the presence of any binary stellar companion of GJ\,3090 up to $\sim 3$ AU as well as any outer planetary companion more massive than 1\,\mjup (resp. 13\,\mjup) within 15\,AU (resp. 100\,AU) assuming co-planarity with GJ\,3090\,b.

The orbit of GJ\,3090\,b, as the innermost planet of the system, is difficult to reconcile with HEM triggered by a massive planetary or stellar wide binary companion. Since this process primarily acts on the outermost planet(s) of the system, it would have either disrupted the inner planet or left it gravitationally attached to the star and thus aligned. In particular, it is difficult for such process to drive GJ\,3090\,b onto a retrograde orbit while preserving the current dynamical stability of the other planet(s) \citep{Lamontagne2026}. Moreover, the companion limits we derived fully exclude the presence of a wide stellar binary and strongly restrict the presence of a massive outer planetary companion, allowing only highly non-coplanar configurations.
GJ\,3090\,b is therefore the first retrograde planet in a multi-planetary system with neither a confirmed massive outer companion nor a wide binary stellar companion. This disfavors late gravitational interactions (such as HEM or secular interaction) as a likely explanation for the strong misalignment observed here and may instead point to a primordial origin for the planet's orbit tilt.
\vspace{-0.3cm}
\vspace{-0.8cm}

\subsection{Formation scenarios}

Some mechanism is therefore needed either to flip the primordial disk or to produce a misaligned secondary disk. Misalignments in disk material are increasingly understood to be widespread, as supported for example by the high incidence of scattered light shadows in protoplanetary disks \citep{Villenave2023}. A hidden, highly inclined companion (with respect to the primordial disk) at a few astronomical units could cause precession of the primordially aligned disk, similar to the scenario suggested for the K2-290 system \citep{Zanazzi2018,Hjorth2021}. However, the origin of a highly misaligned planetary companion is difficult to explain early in the disk evolution, while a close stellar binary would truncate the disk, leaving an insufficient amount of solid material to form the planets. The recently discovered system TOI-6963 \citep{Barber2024} hosting both a transiting planet and a face-on protoplanetary disk shows that strong misalignements between planets and disk material is physically possible. Star-disk encounters during the dynamical decay of a triple system \citep{Nealon2025} or the formation of a new disk via late-stage infall of gas from the interstellar medium \citep{Huhn2025} have been suggested as possible drivers of this system's misalignment. 

In the case of GJ\,3090, a stellar encounter is unlikely to produce a retrograde disk, partly because the required retrograde stellar encounter is inefficient at exchanging angular momentum between the disk and the star \citep{Winter2018, Cuello2019}. As a result, late-stage infall appears to be the only viable mechanism to produce the observed misalignment of the planet. Such infall would have a randomly oriented angular momentum vector and may be fairly common based on both observational evidence  \citep[e.g.,][]{Garufi2026} and theoretical predictions \citep[e.g.,][]{Kuffmeier2023, Winter2024}. However, late-stage infall on the primordial disk is unlikely to flip it into retrograde motion for M dwarf hosts \citep{Kuffmeier2024}. The retrograde orbit of GJ\,3090\,b is thus among the strongest evidence to date of a completely replenished second-generation disk, possibly an evolved-stage analog of the so-called Peter Pan disks around M stars, which sustain strong accretion at ages of several tens of megayears \citep{Silverberg2020}. Furthermore, the unique system architecture appears to have made tidal realignment inefficient, despite the expectation that M dwarfs should efficiently dampen misaligned obliquities \citep{Winn2010,Attia2023}. This trend is distinct from the population of close-in giant planets orbiting M dwarfs, which are preferentially well aligned \citep{Weisserman2025}, thus suggesting a distinct formation pathway between hot Jupiters around M dwarfs and GJ\,3090-like systems.

If GJ\,3090 acquired a new retrograde disk early on, we can ask how much mass ($\Delta M_\mathrm{acc}$) could be accreted before the stellar spin axis realigns with that disk. Mathematical details of the following estimates are provided in Appendix~\ref{appdx:disk_realignment}. Comparing the stellar angular momentum to angular momentum carried by the accreted material, the misalignment persists as long as $\Delta M_\mathrm{acc} < 1\,500\,\me$. Assuming a dust-to-gas ratio of $10^{-2}$, this corresponds to a maximum dust mass of $\sim 15\,\me$. 
Given that the inner planets of the system constitute a significant fraction of the maximum dust mass accreted, most of the dust mass in the secondary disk formed planets rather than accreted onto the star. This implies either suppressed accretion or highly efficient core formation in the inner disk.
\vspace{-0.3cm}
\vspace{-0.8cm}

\section{Summary}
In this letter, we have studied the orbital architecture of the GJ\,3090 multi-planetary system with NIRPS. We find that GJ\,3090\,b is the first known retrograde planet orbiting an M dwarf, excluding polar orbits ($\psi=90^\circ$) at 2.5\,$\sigma$. With a 3D obliquity of $\psi=136^{+24}_{-18}\,\mathrm{deg}$, this sub-Neptune is among the most highly misaligned known planets. We further find no evidence of either a massive outer planetary companion or a wide stellar binary companion. In contrast to the five other confirmed multi-planetary systems hosting a highly misaligned planet ($\psi>70^\circ$), GJ\,3090 is the first system found to lack both. 

Whereas highly misaligned orbits are commonly attributed to gravitational interactions with a massive companion, the architecture of the GJ\,3090 system instead favors a primordial misalignment of the protoplanetary disk. In particular, the retrograde orbit of GJ\,3090\,b points toward a late-stage infall during which the stellar host accreted a secondary disk that was not aligned with the stellar spin axis. Our proposed scenario implies that all planets in the system formed in the same protoplanetary disk then migrated, inheriting its high misalignment and remaining mutually coplanar.

\bibliographystyle{aa} 
\bibliography{bib}

\clearpage

\begin{appendix}

\section{Acknowledgements}
\begin{acknowledgements}
This project has received funding from the European Research Council (ERC) under the European Union's Horizon 2020 research and innovation programme (project {\sc Spice Dune}, grant agreement No 947634). This material reflects only the authors' views and the Commission is not liable for any use that may be made of the information contained therein.
This work has been carried out within the framework of the NCCR PlanetS supported by the Swiss National Science Foundation under grants 51NF40\_182901 and 51NF40\_205606.
AJW has been supported by the Royal Society through a University Research Fellowship grant number URF\textbackslash R1\textbackslash 241791.
RAl acknowledges the Swiss National Science Foundation (SNSF) support under the Post-Doc Mobility grant P500PT\_222212 and the support of the Institut Trottier de Recherche sur les Exoplan\`etes (IREx).
RAl, CC, PL, \'EA, FBa, BB, NJC, RD, LMa \& JPW acknowledge the financial support of the FRQ-NT through the Centre de recherche en astrophysique du Qu\'ebec as well as the support from the Trottier Family Foundation and the Trottier Institute for Research on Exoplanets.
Research activities of the Board of Observational and Instrumental Astronomy at the Federal University of Rio Grande do Norte (NAOS) are supported by continuous grants from the Brazilian funding agency CNPq. This study was financed in part by the Coordena\c{c}\~ao de Aperfei\c{c}oamento de Pessoal de N\'ivel Superior -- Brasil (CAPES) -- Finance Code 001, and by the program CAPES/Print.
ICL acknowledges CNPq research fellowships (Grant No. 313103/2022-4).
VMP, JIGH, RRe, ASM \& AKS acknowledge financial support from the Spanish Ministry of Science, Innovation and Universities (MICIU) projects PID2020-117493GB-I00 and PID2023-149982NB-I00.
PL acknowledges financial support from the Severo Ochoa grant CEX2021-001131-S funded by MCIN/AEI/10.13039/501100011033.
PL is funded by the European Union (ERC, THIRSTEE, 101164189). Views and opinions expressed are however those of the author(s) only and do not necessarily reflect those of the European Union or the European Research Council. Neither the European Union nor the granting authority can be held responsible for them."
\'EA, FBa, RD \& LMa acknowledges support from Canada Foundation for Innovation (CFI) program, the Universit\'e de Montr\'eal and Universit\'e Laval, the Canada Economic Development (CED) program and the Ministere of Economy, Innovation and Energy (MEIE).
SCB, EC \& NCS acknowledge the support from FCT - Funda\c{c}\~ao para a Ci\^encia e a Tecnologia through national funds by these grants: UIDB/04434/2020, UIDP/04434/2020.
SCB acknowledges the support from Funda\c{c}\~ao para a Ci\^encia e Tecnologia (FCT) in the form of a work contract through the Scientific Employment Incentive program with reference 2023.06687.CEECIND and DOI \href{https://doi.org/10.54499/2023.06687.CEECIND/CP2839/CT0002}{10.54499/2023.06687.CEECIND/CP2839/CT0002.}
XB, XDe \& TF acknowledge funding from the French ANR under contract number ANR\-24\-CE49\-3397 (ORVET), and the French National Research Agency in the framework of the Investissements d'Avenir program (ANR-15-IDEX-02), through the funding of the ``Origin of Life" project of the Grenoble-Alpes University.
BLCM acknowledge CAPES postdoctoral fellowships.
BLCM acknowledges CNPq research fellowships (Grant No. 305804/2022-7) and Universal (grant no. 408100/2025-7).
RC acknowledges support from the Canada Research Chairs Program and the Natural Sciences and Engineering Research Council of Canada (NSERC).
NBC acknowledges support from an NSERC Discovery Grant, a Canada Research Chair, and an Arthur B. McDonald Fellowship, and thanks the Trottier Space Institute for its financial support and dynamic intellectual environment.
JRM acknowledges CNPq research fellowships (Grant No. 308928/2019-9).
XDu acknowledges the support from the European Research Council (ERC) under the European Union’s Horizon 2020 research and innovation programme (grant agreement SCORE No 851555) and from the Swiss National Science Foundation under the grant SPECTRE (No 200021\_215200).
DE acknowledge support from the Swiss National Science Foundation for project 200021\_200726. The authors acknowledge the financial support of the SNSF.
CMo acknowledges the funding from the Swiss National Science Foundation under grant 200021\_204847 “PlanetsInTime”.
Co-funded by the European Union (ERC, FIERCE, 101052347). Views and opinions expressed are however those of the author(s) only and do not necessarily reflect those of the European Union or the European Research Council. Neither the European Union nor the granting authority can be held responsible for them.
ASM acknowledges financial support from the Spanish Ministry of Science and Innovation (MICINN) under the 2024 Ram\'on y Cajal program MICINN RYC2024-050707-I.
GAW is supported by a Discovery Grant from the Natural Sciences and Engineering Research Council (NSERC) of Canada.
KAM acknowledges support from the Swiss National Science Foundation (SNSF) under the Postdoc Mobility grant P500PT\_230225.
ED acknowledges support from a Banting Postdoctoral Fellowship - NSERC.
LD \& ED acknowledge financial support from the Faculty of Science at the University of Waterloo, and support from the Waterloo Centre for Astrophysics.
Y.G.C.F. acknowledges funding from the Swiss National Science Foundation (SNSF) through project P500PN\_217951.
AKS acknowledges financial support from La Caixa Foundation (ID 100010434) under the grant LCF/BQ/DI23/11990071.

\end{acknowledgements}

\section{Data reduction}
\label{appdx:reduction_steps}

\subsection{Individual transit observations}

Observational details for each visit are summarized in Table~\ref{tab:observations}. The transit duration from first to fourth contact is $\sim$1.25 hours, and the total observed sequence was scheduled to ensure a summed pre- and post-transit baseline exceeding one hour, though the exact duration varies between visits. For all six visits, we selected the high-efficiency (HE) mode of NIRPS, which reduces the impact of modal noise at the reddest wavelengths compared to the high-accuracy (HA) mode. We note that the first two transits differ from the usually adopted high-accuracy (HAM) mode for HARPS in combined transit observations. Moreover, due to technical issues with HARPS, two visits could not benefit from the combination of both instruments. The S/N remains consistently stable across the six visits, owing to the assistance of the NIRPS adaptive optics (AO) loop. We note, however, that during the second part of the 4 September 2024 transit, the AO loop opened, resulting in the loss of three exposures. Further instrumental issues on that night prevented observations of the post-transit baseline. Finally, given the much lower brightness of GJ\,3090 in the visible compared to near-infrared, HARPS exposure times were increased. 
\vspace{-0.3cm}

\subsection{Reduction steps}

Reduction steps applied in \texttt{ANTARESS} are described in details in \citet{Bourrier2024}. In this section we only summarize the main steps we followed: 

\begin{enumerate}
    \item Raw 2D echelle spectra: extracted using the DRS of NIRPS v3.3.12 and HARPS v3.3.6, both adapted from the ESPRESSO pipeline \citep{Pepe2021}. 
    \item Telluric absorption correction: is based on \citet{Allart2022} and refined over the HARPS and NIRPS observations independently. By computing the CCF of the tellurics in our observed spectrum, we optimized the line list of methane and water telluric lines over which the absorption profile is fit (pressure, temperature, and column density) by minimizing the residual CCF dispersion in individual exposures. 
    \item Telluric emission correction: near-infrared observations further suffer from heavy OH telluric emission contamination. We used the NIRPS DRS emission correction, described in \citet{Srivastava2026}, that we propagated into the \texttt{ANTARESS} pipeline. 
    \item Flux balance correction of Earth diffusion.
    \item Cosmics correction and bad-pixel removing.
    \item CCF mask computation: see Sect.~\ref{appdx:CCF_mask}.
    \item Detrending: computed with the custom CCF mask. Individually for each visit and instrument, we extracted from the disk-integrated stellar spectra the time-series of CCF profiles. We then fit each exposure with a line profile to derive parameters of the averaged stellar line. We note that we used a double Gaussian profile to reproduce the Gaussian core and side lobes, often observed in M dwarf CCFs \citep{Bourrier2018}. As shown in Fig.~\ref{fig:classical_RM}, we do not reach a sufficient precision on the residual RVs to identify trends in the measured RV. However, the fit contrast of the CCF was sufficiently constrained such that contrast detrending was applied in each visit on HARPS and NIRPS. 
    \item Alignment in stellar rest frame: based on the systemic velocity derived for each visit with the custom CCF mask and the Keplerian motion. 
    \item Broadband transit flux scaling.
    \item Differential profile extraction between the master out and each exposure echelle spectra.
    \item Intrinsic profile extraction, through normalization of in-transit differential echelle spectra.
    \item Computation of intrinsic CCFs using the custom CCF mask. 
\end{enumerate}

\subsection{CCF mask refinement}
\label{appdx:CCF_mask}
The CCF mask used throughout our analysis was derived using the out-of-transit observations on GJ\,3090, following two different procedures. Firstly, we used the mask computation procedure implemented in \texttt{ANTARESS} and described in details in \citet{Bourrier2024}. The line identification relies on \citet{Cretignier2020}, after which a series of selection steps is applied either on the normalized average stellar spectrum (AS) or the time-series of individual spectra (IS):
\begin{enumerate}
    \item S/N on line integrated contrast (AS).
    \item Line integrated contrast temporal dispersion (IS).
    \item Line depth (AS).
    \item Line width (AS).
    \item Fit line position with respect to the minimum of the line core (AS).
    \item Line depth ratio with telluric lines crossing stellar lines during the observing sequence (IS).
    \item Line morphology and asymmetry (AS).
    \item Fit line RV temporal dispersion (IS).
\end{enumerate}
Secondly, we built another mask using GJ\,3090 time-series published in \citet{Lamontagne2026}, which comprises 115 HARPS and 84 NIRPS exposures. We computed line-by-line RVs using the template-matching method of \citet{Artigau_2022}, implemented in the open-source \texttt{\href{https://github.com/njcuk9999/lbl}{LBL}} package. \texttt{LBL} first builds a co-added template spectrum of all individual exposures and then divides it into "chunks" defined as the spectral regions bounded by two consecutive local maxima. A "chunk" was considered as a spectral line when its depth-to-noise ratio exceeded 5. For each selected line, \texttt{LBL} returns the RV time-series, to which we applied several selection steps. We only retained lines with valid RV measurements in at least 80\% of the observed epochs, thereby excluding lines that frequently fall outside spectral orders at most barycentric Earth RV (BERV). We further removed lines whose RVs show a Pearson correlation with the BERV higher than 0.3, as these are likely contaminated by tellurics. Finally, we removed blended or asymmetric lines and discarded shallow lines with depths below 5\%.

Due to the lower S/N of the HARPS transit data and the absence of two visits, the \texttt{ANTARESS} HARPS custom mask did not improve the precision of the disk-integrated RVs. Comparing \texttt{ANTARESS} and \texttt{LBL} masks on the precision of disk-integrated RVs, we selected the \texttt{ANTARESS} mask for NIRPS and \texttt{LBL} mask for HARPS. For both the NIRPS and HARPS custom masks, each selected line was assigned a CCF weight equal to the inverse square of its median per-line RV uncertainty, estimated using the \citet{Bouchy_2001} formalism. The selected lines are further inspected for strong temporal outliers, using an iterative 5$\sigma$-clipping. Flagged unstable lines were down-weighted using their dispersion instead of nominal RV error in the weight calculation. After applying all selection criteria, the final NIRPS and HARPS CCF masks contained 862 and 7\,460 lines, respectively.

To assess the quality of the custom CCF masks, derived on the observed data, compared to the default M2 DRS mask, we computed and fit the disk-integrated CCFs. Then, we compared their averaged properties and deviation across the time-series for each visit. The results, listed in Table~\ref{tab:drs_mask}, show significant improvement in both the precision and stability of the custom CCF masks over the DRS one.

\begin{table*}[h]
\caption{Six transit observations with NIRPS and HARPS.}
\centering
\small
\begin{tabular}{ll|cccccc}
\hline
        & Transit date & 2023/10/09 & 2023/10/12 & 2023/11/21 & 2024/07/26 & 2024/09/04 & 2025/07/09\\ 
\hline
\multirow{4}{*}{\rotatebox{90}{\textbf{NIRPS}}}
 & Mode              & \multicolumn{6}{c}{HE} \\
 & $\langle$ S/N $\rangle$ & $88_{-6}^{+4}$ & $96_{-6}^{+2}$ & $97_{-5}^{+5}$ & $78_{-10}^{+9}$ & $93_{-1}^{+1}$ & $136_{-6}^{+2}$ \\
 & $N_{\mathrm{In}}/N_{\mathrm{Out}}$            & 15/11 & 14/12 & 15/12 & 15/12 & 11/7 & 8/10 \\
 & $t_\mathrm{exp}$ [s]  & 300  & 300  & 300 & 300 & 300 & 600 \\
\hline
\hline
\multirow{4}{*}{\rotatebox{90}{\textbf{HARPS}}}
 & Mode              & EGGS & EGGS & \diagbox{}{} & \diagbox{}{} & HAM & HAM \\
 & $\langle$ S/N $\rangle$ & $17_{-1}^{+1}$ & $22_{-2}^{+1}$ & \diagbox{}{} & \diagbox{}{} & $16_{-1}^{+1}$ & $26_{-1}^{+1}$ \\
 & $N_{\mathrm{In}}/N_{\mathrm{Out}}$            & 6/6 & 7/5 & \diagbox{}{} & \diagbox{}{} & 8/6 & 5/7 \\
 & $t_\mathrm{exp}$ [s]  & 600  & 600  & \diagbox{}{} & \diagbox{}{} & 600 & 900 \\
\end{tabular}
\tablefoot{While NIRPS observations are only performed using the high-efficiency mode ($\mathcal{R}\sim 75\,000$), we used both high-efficiency (EGGS, $\mathcal{R}\sim 80\,000$) and high-accuracy (HAM, $\mathcal{R}\sim 115\,000$) modes with HARPS depending on the visit. $N_{\mathrm{In}}$ and $N_{\mathrm{Out}}$ indicate the number of in-transit and out-of-transit exposures. We report the 16$^{\mathrm{th}}$, 50$^{\mathrm{th}}$ and 84$^{\mathrm{th}}$ percentile of S/N measured for in-transit exposures in order 57 (resp. 49) of NIRPS (resp. HARPS). }
\label{tab:observations}
\end{table*}

\begin{table*}[h]
\caption{Stability and precision comparison between the DRS default M2 mask and the custom one derived from GJ\,3090 observations.}
\centering
\small
\begin{tabular}{ll|cccccccccccc}
\hline
        & Transit date & \multicolumn{2}{c}{2023/10/09} & \multicolumn{2}{c}{2023/10/12} & \multicolumn{2}{c}{2023/11/21} & \multicolumn{2}{c}{2024/07/26} & \multicolumn{2}{c}{2024/09/04} & \multicolumn{2}{c}{2025/07/09} \\ 
        &  & DRS & Custom & DRS & Custom & DRS & Custom & DRS & Custom & DRS & Custom & DRS & Custom \\ 
\hline
\multirow{5}{*}{\rotatebox{90}{\textbf{NIRPS}}}
 & $\langle \mathrm{Contrast} \rangle \, [\%]$            & \textbf{30.56} & 28.48 & \textbf{30.46} & 28.39 & \textbf{30.51} & 28.42 & \textbf{30.39} & 28.27 & \textbf{30.55} & 28.41 & \textbf{30.42} & 28.22 \\
 & $\sigma_\mathrm{CTRST}^{rel}$\,[ppm]   & 324 & \textbf{168} & 274 & \textbf{262} & 220 & \textbf{208} & 439 & \textbf{270} & \textbf{135} & 164 & 145 & \textbf{133} \\
 & $\langle \mathrm{FWHM} \rangle \, [\kms]$ & 7.16 & \textbf{6.65} & 7.21 & \textbf{6.65} & 7.18 & \textbf{6.64} & 7.18 & \textbf{6.61} & 7.19 & \textbf{6.65} & 7.20 & \textbf{6.62} \\
 & $\sigma_\mathrm{FWHM}^{rel}$\,[ppm] & 369 & \textbf{242} & 320 & \textbf{239} & 369 & \textbf{285} & 508 & \textbf{252} & 236 & \textbf{185} & 201 & \textbf{148} \\
 & $\sigma_\mathrm{RVres}$\,[\ms]  & 9.81 & \textbf{7.81} & 7.01 & \textbf{5.26} & 8.51 & \textbf{6.09} & 11.20 & \textbf{9.26} & 6.88 & \textbf{6.80} & 3.79 & \textbf{3.33} \\
\hline
\hline
\multirow{5}{*}{\rotatebox{90}{\textbf{HARPS}}}
 & $\langle \mathrm{Contrast} \rangle \, [\%]$    & 17.36 & \textbf{17.61} & 17.36 & \textbf{17.61} & \diagbox{}{} & \diagbox{}{} & \diagbox{}{} & \diagbox{}{} & \textbf{22.78} & 21.68 & \textbf{23.50} & 22.15 \\
 & $\sigma_\mathrm{CTRST}^{rel}$\,[ppm]  & 423 & \textbf{300} & 237 & \textbf{148} & \diagbox{}{} & \diagbox{}{} & \diagbox{}{} & \diagbox{}{} & 449 & \textbf{380} & 277 & \textbf{95} \\
 & $\langle \mathrm{FWHM} \rangle \, [\kms]$ & \textbf{5.45} & 5.67 & \textbf{5.50} & 5.66 & \diagbox{}{} & \diagbox{}{} & \diagbox{}{} & \diagbox{}{} & \textbf{4.69} & 5.10 & \textbf{4.61} & 4.92 \\
 & $\sigma_\mathrm{FWHM}^{rel}$\,[ppm]  & 444 & \textbf{279} & 278 & \textbf{125} & \diagbox{}{} & \diagbox{}{} & \diagbox{}{} & \diagbox{}{} & 340 & \textbf{299} & 234 & \textbf{124} \\
 & $\sigma_\mathrm{RVres}$\,[\ms]  & 9.74 & \textbf{7.13} & \textbf{5.00} & 5.62 & \diagbox{}{} & \diagbox{}{} & \diagbox{}{} & \diagbox{}{} & 12.85 & \textbf{2.79} & 3.55 & \textbf{3.41} \\
\end{tabular}
\tablefoot{The $rel$ subscript indicates that the value is computed with respect to the mean over out-of-transit exposures. Compared best values between the DRS and custom masks are reported in bold.}
\label{tab:drs_mask}
\end{table*}

\section{Rossiter-McLaughlin fit}
\label{appdx:RM_fit}

\begin{figure}[ht]
    \centering
    \includegraphics[width=0.7\columnwidth]{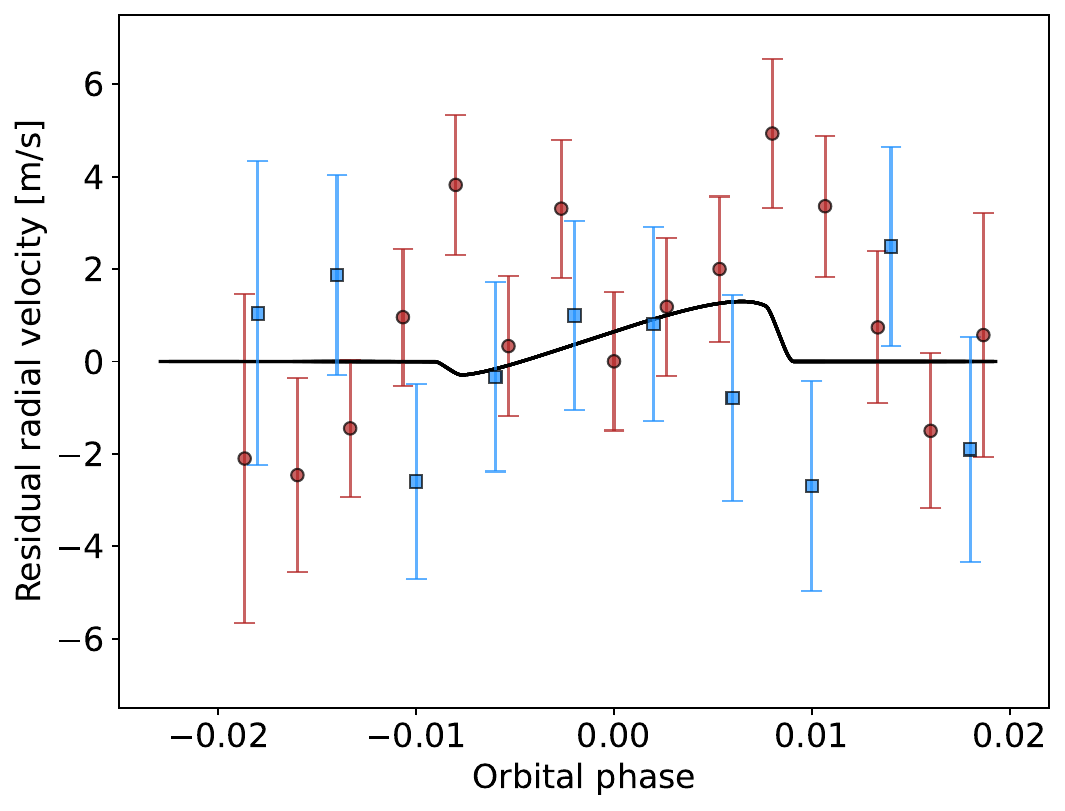}
    \caption{Measured residual RVs from the Keplerian. The black line denotes the predicted RM anomaly of our best-fit model. The binned NIRPS (resp. HARPS) RVs from the six (resp. four) visits are represented by the red circles (resp. blue squares) markers.}
    \label{fig:classical_RM}
\end{figure}
\vspace{-0.2cm}
The RM effect is modeled using the RM Revolutions technique, described in detail in \citet{Bourrier2021}. We note at this stage that the classical RM did not yield any constraints on the obliquity of GJ\,3090\,b, as highlighted in Fig.~\ref{fig:classical_RM} where the RVs scatter is much larger than amplitude of the RM anomaly. In the RMR approach, the local stellar spectra occulted by the planet (intrinsic profiles) are extracted from the disk integrated observed time-series of flux spectra. The CCFs are then computed over these intrinsic spectra rather than the disk-integrated ones, as is usually done in the classical RM approach. Finally, the time-series of intrinsic CCFs is fit conjointly with a Gaussian line profile to model the local average stellar line occulted by the planet.

\vspace{-0.3cm}
\begin{figure}[ht]
    \centering
    \includegraphics[width=0.7\columnwidth]{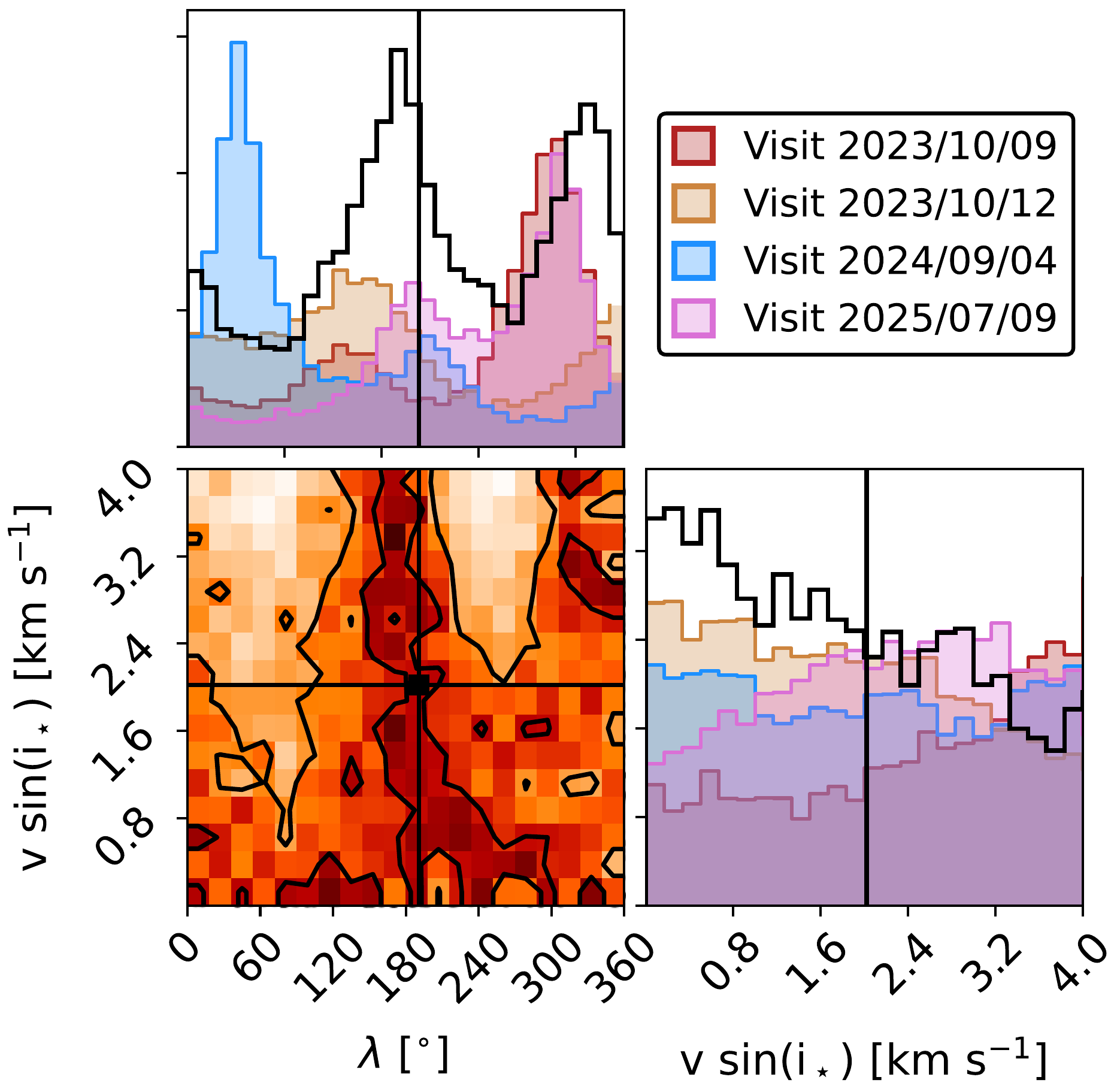}
    \caption{Same as Fig.~\ref{fig:lambda_veq} but for HARPS.}
    \label{fig:lambda_veq_HARPS}
\end{figure}
\vspace{-0.2cm}

\begin{figure*}[ht]
    \centering
    \includegraphics[width=1.2\columnwidth]{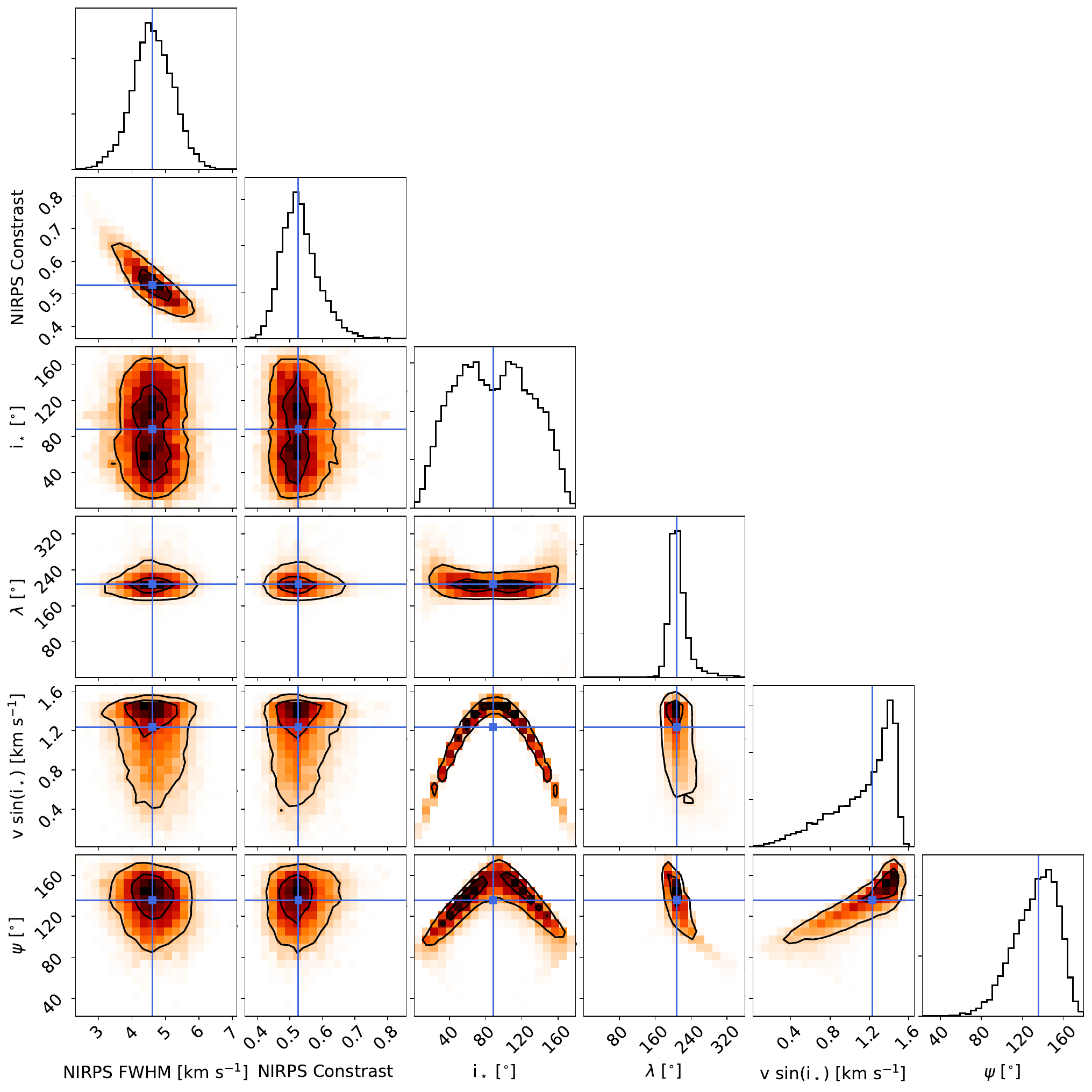}
    \caption{Correlation diagrams for the PDFs of the RMR model parameters for the GJ\,3090\,b combining the five selected NIRPS transits. Black curves show the 1 and $2\sigma$ 2D confidence regions. The blue lines and dots represent the median values of each distribution.}
    \label{fig:full_corner_psi}
\end{figure*}

The intrinsic CCFs are firstly fit independently for the FWHM, contrast, and RV centroid using an MCMC approach. This preliminary step serves to identify ill-defined CCFs, typically associated to the lowest S/N, whose inclusion could bias or increase the noise in the following RMR fits. Based on the posterior distributions of the RVs and FWHM, we retained only exposures that converged in either of these parameters for the final RMR fit, referred thereafter as the quality-selected subset exposures. 

As a next step, we fit the time-series of quality-selected exposures for each visit and instrument independently. The free parameters are simply the projected obliquity $\lambda$ and the projected rotational velocity $v\sin i_*$. This parametrization is the most robust one, as it relies only on the RM signal itself without requiring  priors from the literature. As discussed in Sect.~\ref{sec:RM_analysis}, this per-visit fit yields poorly constrained posteriors for HARPS but well-defined PDFs for NIRPS. This discrepancy is not surprising as NIRPS benefits from both a better temporal sampling and a higher S/N per individual exposure compared to HARPS. Nonetheless, we note that the RMR fit for the NIRPS visit 4 did not converge to an obliquity solution fully consistent with the other five visits, and in particular favored an inconsistently large stellar rotational velocity. To further assess the reliability of each individual fit, we compared the BIC of each best-fit model against that of the null hypothesis (i.e., flat intrinsic CCFs). None of the HARPS visits showed preference for the model over the null, while all NIRPS visits, except for visit 4, strongly favored the fit model with $\Delta \mathrm{BIC}>10$ \citep{Thorngren2026}. We attribute this to the low S/N of the HARPS observations combined with higher exposure times that do not allow a proper sampling of the photospheric RV field. Based on the BIC, we therefore excluded all HARPS visits and the NIRPS visit 4 from the final RMR fit.

In the final step, we performed a joint fit of the NIRPS quality-selected intrinsic CCFs (excluding HARPS and the NIRPS visit 4), adopting a Gaussian profile, and incorporating prior constraints on stellar and planetary parameters from the literature. A summary of all parameters used and derived in this final RMR fit is provided in Table~\ref{tab:RM_fits}, and the posterior distributions of various fit parameters are shown in Fig.~\ref{fig:full_corner_psi}. Because the distribution of $i_*$ is quasi-symmetric around $90^\circ$, the northern and southern configurations of $\psi$ give similar results. The PDFs of $\lambda$, combined $\psi$ and, $v\sin i_*$ are well-defined and nearly entirely exclude polar orbits, pointing toward a retrograde orbit.

\begin{table*}[h]
\caption{Summary of parameters used by or derived from the Rossiter-McLaughlin fit.  }
\renewcommand{\arraystretch}{1.2}
\centering
\small
\begin{tabular}{cccc}
    \hline
        Parameter & Unit & Prior & Retrieved value \\ \hline
    \multicolumn{4}{c}{Planetary parameters}\\ \hline
        Scaled radius ($\rplanet/\rstar$)&  & 0.03853 & \dots \\
        Scaled semi-major axis ($a/\rstar$)& & $13.1686^{+0.1250}_{-0.1250}$ & \dots \\
        Orbital inclination ($i_\mathrm{pl}$)& deg & $86.93^{+0.18}_{-0.16}$ & \dots \\
        Impact parameter ($b$)& & $0.70^{+0.02}_{-0.02}$ & \dots \\
        Eccentricity ($e$)& & 0.0 & \dots \\
        Projected obliquity ($\lambda$)& deg & $\mathcal{U}(-2\pi,2\pi)$ & $208^{+15}_{-18}$ \\
        South 3D obliquity ($\psi_S$)& deg & \dots & $138^{+23}_{-18}$ \\ 
        North 3D obliquity ($\psi_N$)& deg & \dots & $134^{+24}_{-19}$ \\ 
        Combined 3D obliquity ($\psi$)& deg & \dots & $136^{+24}_{-18}$ \\ \hline
        \multicolumn{4}{c}{Stellar parameters}\\ \hline
        
        Radius ($\rstar$)& \rsol & $\mathcal{N}(0.516,0.016)$ & $0.518^{+0.015}_{-0.015}$\\
        Rotation period ($P_\mathrm{rot}$)& days & $\mathcal{N}(17.95,0.15)$ & $17.90^{+0.15}_{-0.15}$\\
        South stellar inclination ($i_{*,S}$)& deg & \dots & $122^{+16}_{-29}$\\
        North stellar inclination ($i_{*,N}$)& deg & \dots & $58^{+29}_{-16}$\\
        Combined stellar inclination ($i_*$)& deg & \dots & $88^{+40}_{-50}$\\
        Quadratic LD ($u_1$, $u_2$)& & 0.201, 0.416 & \dots \\
        Rotational velocity ($v\sin i_*$)& \kms & \dots & $1.23^{+0.28}_{-0.23}$\\

\end{tabular}
\tablefoot{Parameters with no Retrieved value are not directly fit by the model, while those lacking a Prior are reconstructed from other parameters. All prior constraints are drawn from \citet{Lamontagne2026}. }
\label{tab:RM_fits}
\end{table*}

To further demonstrate the robustness of our results, we applied the same fitting procedure to different selections of visits, exposures, and instruments. We defined 3 more sub-datasets as: the full unfiltered dataset, the full unfiltered dataset excluding the NIRPS visit 4, and the quality-selected subset exposures excluding the NIRPS visit 4, both of which include all HARPS exposures. The derived values of $\psi$ for each selection, listed in Table~\ref{tab:RM_compa}, are mutually consistent, confirming that the retrograde orbit solution does not depend on the choice of fit exposures. The quality-selected subset excluding NIRPS visit 4 and HARPS yields the smallest $1\sigma$ error-bars and also produced the best-behaved posteriors for most of the fit parameters. We therefore adopted this sub-selection as our reference dataset for all results reported in this letter. Consistent with the per-visit analysis of the previous steps, the FWHM and contrast posteriors confirmed that the HARPS data contribution to the joint fit is marginal compared to NIRPS, as they fail to converge. Our results are therefore mostly driven by the NIRPS data, which dominate the constraints on the RM signal. Furthermore, we note that we did not take into account stellar activity (spots) in our RM modeling as we had no indication of contamination from the intrinsic CCF maps. Nevertheless, this should not strongly bias the measured obliquity as it is driven from near-infrared data \citep{Oshagh2016}. Finally, \citet{Lamontagne2026} reported tentative evidence of differential stellar rotation. We therefore refit the quality-selected subset of exposures while leaving the differential rotation  as a free parameter. We found that the data do not constrain the stellar differential rotation, and its inclusion has no significant impact on the inferred architecture reported in Table~\ref{tab:RM_fits}.

\begin{table}[h]
\caption{Comparison of the derived 3D obliquity for different exposure selections.}
\renewcommand{\arraystretch}{1.2}
\centering
\small
\begin{tabular}{l|cc}
\hline
        & Derived $\psi$\,[deg]  \\ 
\hline
All exp. \& all vis. \& N + H & $110^{+44}_{-30}$  \\
All exp. \& no vis. 4 \& N + H & $136^{+24}_{-20}$ \\
Sub-exp. \& no vis. 4 \& N + H & $136^{+24}_{-18}$ \\
Sub-exp. \& no vis. 4 \& N only & $136^{+24}_{-18}$ \\
\end{tabular}
\tablefoot{ N and H refer, respectively, to NIRPS and HARPS. The fourth visit corresponds to the 26 July 2024 transit.}
\label{tab:RM_compa}
\end{table}

\section{System characterization}
\label{appdx:system_properties}

\begin{figure}[ht]
    \centering
    \includegraphics[width=0.85\columnwidth]{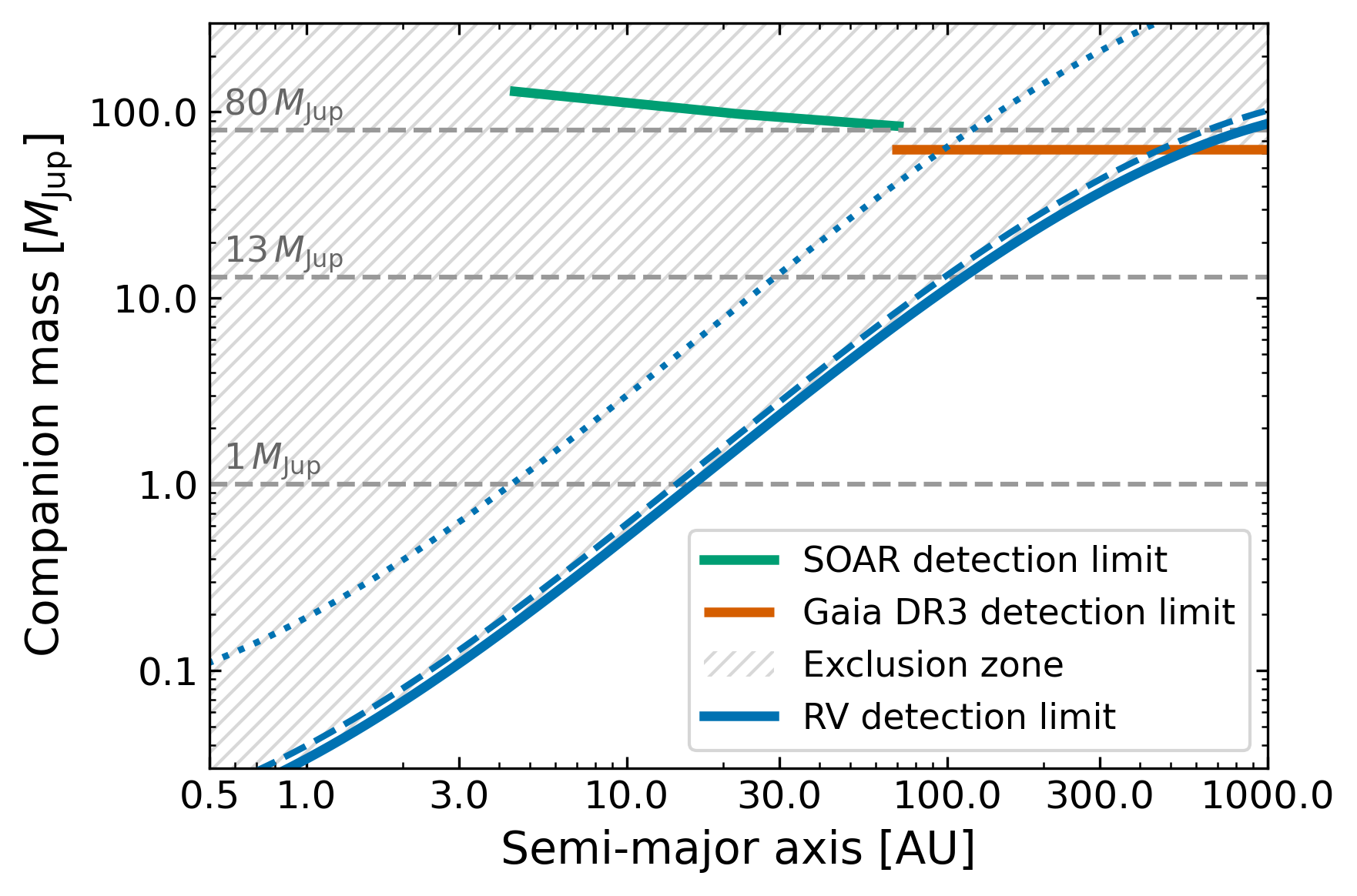}
    \vspace{-0.2cm}
    \caption{Detection limits on companion mass as a function of semi-major axis for GJ\,3090. The blue, green, and orange curves correspond to the RV, SOAR speckle-imaging, and Gaia DR3 constraints, respectively. The three blue curves are obtained assuming different inclinations ($i_c$) for the companion: $i_c=i_\mathrm{pl}$ (solid), $i_c=i_{*,S}$ (dashed) and $i_c=10^\circ$ (dotted). The hatched region denotes the parameter space excluded by the combined observations (coplanar case). Horizontal dashed gray lines at 13~\mjup\ and 80~\mjup\ indicate the approximate transitions between planetary, brown dwarf, and stellar companions.  }
    \label{fig:companion_mass_limit}
\end{figure}

\noindent We combined two age proxies for GJ\,3090. The empirical gyrochronology relation for M dwarfs from \citet{Engle2023}, which infers age from the stellar rotation period alone, yields $1.17^{+0.17}_{-0.15}$\,Gyr. We further used the code \texttt{stardate} \citep{Angus2019b}, which combines isochrone fitting with gyrochronology \citep{Angus2019}, finding an age of $0.81^{+1.57}_{-0.11}$ Gyr. While both approaches rely on gyrochronology, the relation implemented in \texttt{stardate} is mostly independent of the relation from \citet{Engle2023}, as there are built from different datasets. We therefore treated the two proxies as independent, combining them with bootstrapping yields an age for GJ\,3090 of $1.07_{-0.16}^{+0.71}\,\mathrm{Gyr}$.

The additional RV data we used to constrain outer companions consist of 84 NIRPS and 115 HARPS measurements. The HARPS data collected on GJ\,3090, which span a significantly longer baseline than the NIRPS observations, exhibit evidence of a signal with a period of $\sim$ 3\,300 days. A detailed analysis performed in \citet{Lamontagne2026}, including a comparison with long-term ASAS-SN photometry, strongly suggests that this signal is associated with the stellar magnetic cycle rather than with an orbiting companion. We therefore used the RVs residuals obtained after subtraction of the full model from \citet{Lamontagne2026} (including the two confirmed planets, the planetary candidate, the magnetic cycle, and the activity modeling) to determine which regions of the companion mass--orbital separation parameter space remain compatible with the data.

To derive the RV detection limits, we used the \texttt{kima} package \citep{Faria2018}. We performed a grid search including up to four Keplerian signals, allowing for eccentric orbits, and using periods ranging from 100 to $10^8$ days and semi-amplitudes from $10^{-5}$ to $10^{5}\,\ms$. The fit was performed with $10^6$ nested sampling iterations to ensure adequate exploration of the parameter space. As expected, no additional companion was significantly detected ($>5\sigma$). However, the posterior samples provide information on which regions of parameter space cannot fully exclude the presence of a companion. We therefore used the detection-limit routine implemented in \texttt{kima}, based on the compatibility-map approach described in \citet{Faria2025}. The posterior samples are projected onto the mass--semi-major axis plane, where the highest compatible masses are identified as a function of separation. The resulting distribution is then binned and smoothed to produce a continuous detection-limit curve, shown as the blue line in Fig.~\ref{fig:companion_mass_limit}. 

Since RV observations can only constrain $M \sin(i_\mathrm{pl})$, the inferred mass limits depend on the orbital inclination of the companion ($i_c$). We therefore computed detection limits for several inclinations. The nominal case assumes co-planarity with GJ\,3090\,b, adopting its measured inclination $i_c = i_\mathrm{pl}$, while additional limits were derived for $i_c = i_*$ and $i_c = 10^\circ$. From the coplanar case, we can exclude companions more massive than approximately 1~\mjup within $\sim$15 AU, more massive than the brown dwarf threshold (13~\mjup) within $\sim$100 AU, and stellar companions within $\sim$800 AU. As $i_c$ is reduced, the mass limits become less stringent since a larger true companion mass is required to produce the same observed RV semi-amplitude. Nevertheless, the RV detection limits remain informative, as they exclude most massive companion configurations, allowing only orbits oriented close to perpendicular to the line of sight. We note that our RV detection limit is similar to \citet{Lamontagne2026} with an increased range of semi-major axis.

Additionally, we complemented RV constraints with the SOAR speckle-imaging observations presented by \citet{Almenara2022}. Angular separations were converted into projected physical separations using the distance to GJ\,3090, while the contrast limits were converted into companion-mass limits using the stellar mass and $I$-band magnitude of GJ\,3090, together with the conversion table in \citet{Pecaut2013}. The resulting detection limit is shown by the green curve in Fig.~\ref{fig:companion_mass_limit}. These observations primarily probe a region of parameter space that is already excluded by the RV monitoring.

Finally, Gaia DR3 provides additional constraints on wide stellar binary companions. At a distance of only 22.475 pc, Gaia is sensitive to companions located beyond approximately 3 arcsec ($\sim70$ AU), where sources can be detected down to an absolute magnitude of $M_G \simeq 19$ \citep{GaiaDR32023}. Using the table of \citet{Pecaut2013}, this corresponds to a stellar mass of approximately $0.06~M_\odot$. The resulting detection limit is shown by the orange curve in Fig.~\ref{fig:companion_mass_limit}.

\section{Disk realignment}
\label{appdx:disk_realignment}

Consider the stellar angular momentum, given by 
\begin{equation}
    J_{*} = k_{*}^2 \mstar R_{*}^2  \Omega_{*},
\end{equation}
where $k_{*}^2$ is the dimensionless moment-of-inertia coefficient, $\mstar$ is the stellar mass, $\rstar$ is the stellar radius and $\Omega_{*}$ is the stellar rotation frequency. On the other hand, the angular momentum of accreted material is defined as 
\begin{equation}
    J_\mathrm{acc} = \Delta M_\mathrm{acc} \sqrt{G \mstar R_\mathrm{acc}},
\end{equation}
where $\Delta M_\mathrm{acc}$ is the accreted mass and $R_\mathrm{acc}$ the accretion radius. We further consider for simplicity $R_\mathrm{acc}\sim5 \rstar$, $k_*^2=0.2$, and $\Omega_* = 0.1\,\Omega_{\rm crit}$ using $\Omega_{\rm crit}=\sqrt{{G M_{*}}/{R_{*}^3}}$. In this case the stellar spin-axis is not realigned with the disk as long as $J_\mathrm{acc} \ll J_{*}$ or
\begin{equation}
    \Delta M_\mathrm{acc} \ll \frac{k_{*}^2}{\sqrt{5}} \frac{\Omega_{*}}{\Omega_{\rm crit}} \mstar \sim 1\,500\me.
\end{equation}

\end{appendix}

\end{document}